\documentclass[11pt,a4paper]{article}
\usepackage[utf8]{inputenc}
\usepackage{authblk}
\usepackage{indentfirst}
\usepackage{misccorr}
\usepackage{graphicx}
\usepackage{amsmath}
\usepackage{amssymb}
\usepackage{multicol}
\usepackage{bm}
\usepackage{longtable}
\usepackage{pdflscape}
\usepackage{epstopdf}
\usepackage{multirow}
\usepackage{adjustbox}
\usepackage{amsfonts}
\usepackage{fancyhdr}
\usepackage{ifthen}
\usepackage{fancyheadings}
\usepackage{setspace}
\usepackage[labelfont=bf]{caption}
\newcommand{\ie}{\textit{i.e.,~}}
\newcommand{\eg}{\textit{e.g.,~}}

\fancypagestyle{plain}{\chead{\scriptsize \texttt{Communications of BAO, Vol.2 (LXV), 2018, Is.2, pp.338-345}}}
\title{On the importance of resistivity and Hall effect in MHD simulations of binary neutron star mergers}
\author{A. S. Harutyunyan}
\affil{\emph{\scriptsize NAS RA V. Ambartsumian Byurakan Astrophysical Observatory (BAO), Armenia}\\
\emph{\scriptsize Yerevan State University (YSU), Armenia}\\
\emph{\scriptsize E-mail: arus@bao.sci.am}}

\begin{document}
\pagestyle{empty}
\newpage
\pagestyle{fancy}
\label{firstpage}
\date{}
\maketitle

\begin{abstract}
   We examine the range of rest-mass densities, temperatures and
magnetic fields involved in simulations of binary neutron star mergers (BNSM) and identify the conditions under which the ideal magneto-hydrodynamics (MHD) breaks down using recently computed conductivities of warm, magnetized plasma created in such systems. While
previous dissipative MHD studies of BNSMs assumed that dissipation
sets in due to low conduction at low rest-mass densities, we show that
this paradigm must be shifted: the ideal MHD is applicable up to the
regime where the hydrodynamic description of matter breaks down.
We also find that the Hall effect can be important at low densities and
low temperatures, where it can induce a non-dissipative rearrangement
of the magnetic field. Finally, we mark the region in temperature-density 
plane where the hydrodynamic description breaks down.
\end{abstract}


\section{Introduction}
The recent detections of gravitational waves by the LIGO and Virgo
detectors have opened a new chapter in multimessenger astronomy.
The observations of GW170817 and GRB170817A provided the first direct evidence that a class of short gamma-ray bursts (GRB) can be associated with
the inspiral and merger of binary compact stars. As the magnetic field
plays a central role in the generation of the GRBs, the physics of
inspiral and merger of magnetized neutron stars can now be constrained
directly by observations. The general-relativistic MHD simulations
of these processes have advanced steadily over the recent years (Faber 2012, Paschalidis 2016, Baiotti 2016) and can be carried out either in the case of ideal (Rezzolla 2011, Kiuchi 2014, Palenzuela 2015, Kawamura 2016, Ruiz 2016, Kiuchi 2017, Ruiz 2018) or resistive MHD (Dionysopoulou 2012, Palenzuela 2013b, Palenzuela 2013a, Dionysopoulou 2015).

The aim of this work is to investigate the conditions under which the
non-ideal MHD effects can be important in the contexts of BNSMs and the evolution of the post-merger object.
For doing so, we first discuss briefly the relevant time- and lengthscales over which the physical  parameters evolve in the simulations of
magnetized BNSMs. 
Characteristic timescales of simulations set the
upper bound $\tau_0$ over which the relevant quantities need to change
to be relevant dynamically. The current highest resolution
simulations of the merger process (Siegel 2013, Kiuchi 2015, Kiuchi 2017) are
limited in time after merger due to numerical costs. Lower-resolution
simulations can be carried out up to the point of the collapse to a
black hole or the formation of a differentially rotating and unstable
neutron star, and last typically tens of milliseconds (Dionysopoulou 2012, Dionysopoulou 2015).

To extract the relevant lengthscales we use an ideal MHD simulation of the merger
of an equal-mass magnetized binary system with a total ADM
mass of $M_{\rm ADM}= 3.25 M_{\odot}$ and initial orbital separation of
$45\,{\rm km}$. The finest resolution of the simulation is $l_0 \approx 227\,{\rm m}$, and the merger takes place at $t_{\rm mg} \simeq 3.5\, {\rm ms}$ after the start of the simulation (see ref. Harutyunyan 2018 for details).

The main features of the ideal-MHD simulations
which are important for our estimates are:
\textit{(a)} the characteristic lengthscales over which the magnetic-field
variations can be significant are $\lambda_B\simeq 1\,{\rm km}$ or less, with the lower limit obviously given by the
resolution of the simulation; \textit{(b)} the characteristic
lengthscales over which the rest-mass density variations can be significant are  $\lambda_\rho\simeq 10\,{\rm km}$, \ie the density of matter is approximately constant over
the lengthscales of variation of magnetic field; \textit{(c)} the
characteristic timescales relevant for the simulations are of the
order of $\tau_0\simeq 10\,{\rm ms}$. 

We start this review by collecting the relevant formulae for 
the characteristic timescales which describe the evolution of the magnetic field in Sec. 2. In Sec. 3 we present our numerical results and identify the density-temperature regimes where the ideal MHD approximation as well as MHD description of plasma are justified. Our results are summarized in Sec. 4.

\section{Ohmic diffusion and Hall timescales} 

As is well-known, the MHD
description of low-frequency phenomena in neutron stars is based on
the following Maxwell equations
\begin{eqnarray}
\label{eq:Maxwell_eqs}
   \bm\nabla \times \bm E=-\frac{1}{c}\frac{\partial\bm B}{\partial t},\qquad
   \bm\nabla \times \bm B=\frac{4\pi}{c}{\hat{\sigma}}\bm E,
\end{eqnarray}
where ${\hat{\sigma}}$ is the electrical conductivity tensor. Eliminating the electric field $\bm E$, we obtain an evolution equation for the magnetic-field
\begin{eqnarray}
  \label{eq:mag_field}
  \frac{\partial {\bm B}}{\partial t}=
  -{\bm\nabla \times} \left({\hat{\varrho}} \, {\bm\nabla
    \times}{\bm B}\right),
\end{eqnarray}
where ${\hat{\varrho}}$ is the electrical resistivity tensor defined as 
${\hat{\varrho}} =
({c^2}/{4\pi}){{\hat{\sigma}}}^{-1}$.

In the case of isotropic conduction $\hat\sigma_{ij} =\delta_{ij}\sigma$, and
Eq.~\eqref{eq:mag_field} reduces to
\begin{eqnarray}
  \label{eq:mag_field_is1}
  \frac{4\pi\sigma}{c^2}\frac{\partial \bm B}{\partial t}=
  \Delta\bm B.
\end{eqnarray}
For qualitative estimates we 
approximate $|\Delta \bm B|\simeq B/\lambda_B^2$ and
$|\partial \bm B/\partial t|\simeq B/\tau_d$,
from where we find an estimate for the magnetic field decay (Ohmic diffusion) timescale $\tau_d$
\begin{eqnarray}
  \label{eq:decay_time}
\tau_d = \frac{4\pi\sigma \lambda^2_B}{c^2}.
\end{eqnarray}

In the presence of strong magnetic
fields the electrical conductivity and resistivity tensors can be decomposed as
($b_k=B_k/B$)
\begin{eqnarray}
  \label{eq:sigma_kj}
  \sigma_{kj}&=&\delta_{kj}\sigma_0-\epsilon_{kjm}b_m
  \sigma_1 +b_kb_j\sigma_2,\\
 \label{eq:varrho_kj}
  \varrho_{ik}&=&\delta_{ik}\varrho_0+\epsilon_{ikm}b_m
  \varrho_1 +b_ib_k\varrho_2.
\end{eqnarray}
The components of the resistivity tensor $\hat{\rho}$ are related to those of $\hat{\sigma}$ by 
\begin{eqnarray}
  \label{eq:varrho_0}
  \varrho_0=\frac{c^2}{4\pi}\frac{\sigma_0}{\sigma_0^2+\sigma_1^2},\quad
  \varrho_1=\frac{c^2}{4\pi}\frac{\sigma_1}{\sigma_0^2+\sigma_1^2},\quad
  \varrho_2=\frac{c^2}{4\pi\sigma}\frac{\sigma_1^2-\sigma_0\sigma_2}{\sigma_0^2+\sigma_1^2},
\end{eqnarray}
where $\sigma=\sigma_0+\sigma_2$ is the longitudinal 
conductivity,
\ie the electrical conductivity in the absence of magnetic field. 
The three
components of the conductivity tensor
are given by the following Drude-type formulas (Harutyunyan 2016)
\begin{eqnarray}
\label{eq:sigma_drude}
\sigma = \frac{n_ee^2c^2\tau}{\varepsilon},\quad
\sigma_0 = \frac{\sigma} {1+(\omega_{c}\tau)^2},\quad
\sigma_1 =\frac{(\omega_{c}\tau) \sigma} 
{1+(\omega_{c}\tau)^2}=(\omega_c\tau)\sigma_0.
\end{eqnarray}
 Here $n_e$ is
the electron number density, $e$ is the elementary charge, $c$ is the speed of light, $\tau$ is the electron mean collision time, $\omega_c := ecB\varepsilon^{-1}$ is
the cyclotron frequency, and $\varepsilon$ is the characteristic
energy-scale of electrons.

From Eqs.~\eqref{eq:varrho_0} and 
\eqref{eq:sigma_drude} we obtain the following estimates for the components of the resistivity tensor
\begin{eqnarray}
  \label{eq:varrho_0_Drude}
\varrho_0\simeq \frac{c^2}{4\pi\sigma},\quad
\varrho_1 \simeq (\omega_c\tau)\varrho_0,\quad
  \varrho_2\simeq 0.
\end{eqnarray}
Now, upon using Eqs.~\eqref{eq:varrho_kj} and \eqref{eq:varrho_0_Drude} and approximating again $\partial_i B\simeq B/\lambda_B$, we 
estimate the right-hand side of Eq.~\eqref{eq:mag_field}
\begin{eqnarray}
  \label{eq:A_estimate}
  |{\hat{\varrho}} \, {\bm\nabla \times}\bm B|\simeq 
  {\rm max}(1,\omega_c\tau) \,\varrho_0 \frac{B}{\lambda_B}.
\end{eqnarray}
In the case of small magnetic fields $\omega_c\tau\ll 1$ we
recover the isotropic case from Eqs.~\eqref{eq:mag_field} and \eqref{eq:A_estimate}. The
evolution of magnetic field is then determined by the Ohmic diffusion
timescale $\tau_d$ given by Eq.~\eqref{eq:decay_time}. Conversely, in
the strongly anisotropic regime, where $\omega_c\tau\gg 1$, the
magnetic field evolution is determined by the characteristic timescale
$\tau_B $ given by 
\begin{eqnarray}
\label{eq:B_time}
\tau_B  = \frac{\tau_d}{\omega_c\tau}
=\frac{ 4\pi n_ee \lambda^2_B}{cB}
=\frac{ 4\pi e\rho \lambda^2_B}{cB}\frac{Z}{Am_n}.
\end{eqnarray}
In the last step we used the condition of the charge neutrality, which implies $n_e = Z\rho/(Am_n)$, where $Z$ and $A$ are the charge and the mass number of nuclei, respectively, and $m_n$ is the atomic mass unit. 

Thus, in the strongly anisotropic regime which is realized for
sufficiently high magnetic fields and low rest-mass densities the
characteristic timescale over which the magnetic field evolves is
reduced by a factor of $\omega_c\tau$ due to the Hall effect. We see
from Eq.~\eqref{eq:B_time} that the timescale $\tau_B$ decreases with
an increase of the magnetic field and, in contrast to the Ohmic diffusion timescale $\tau_d$, is independent of the electrical
conductivity $\sigma$. The
physical reason for this difference lies in the fact that the Hall
effect {\it per-se} is not dissipative. Note however that it can act
to facilitate Ohmic dissipation. For instance, the Hall effect may
cause the fragmentation of magnetic field into smaller structures
through the Hall instability, which can then accelerate the decay of
the field via standard Ohmic dissipation (Gourgouliatos 2016, Kitchatinov 2017).

\section{Results}
\label{sec:numerics}

In order to estimate the Ohmic diffusion timescale given by
Eq.~\eqref{eq:decay_time} we use the following fit formula for the
electrical conductivity $\sigma$ (Harutyunyan 2016, Harutyunyan 2017)
\begin{eqnarray}
  \label{eq:sigma_fit}
  \sigma = \frac{1.5\times 10^{22}}{Z}
  \left(\frac{T_F}{1~{\rm MeV}}\right)^a
  \bigg(\frac{T}{T_F}\bigg)^{-b}
  \bigg(\frac{T}{T_F}+d\bigg)^{c}\, {\rm s}^{-1},
\end{eqnarray}
where $T$ and $T_F= 0.511 \big[\sqrt{1+(Z\rho_6/A)^{2/3}}-1\big]\,{\rm MeV}$ are the temperature of stellar matter and the Fermi
temperature of electrons, respectively,
$\rho_6:= \rho/(10^6\,{\rm g\  cm}^{-3})$, and the fitting parameters $a,b,c,d$ are functions of $Z$.

At low-density ($\rho_6 \lesssim 1$) 
and high-temperature ($T\gtrsim 1\,{\rm MeV}$)
regime of stellar matter where we are mainly interested, the matter consists mainly of hydrogen nuclei. In this case $a=0.924,\, b=0.507,\, c=1.295,\, d=0.279$~ (Harutyunyan 2017), and, 
approximating also $T_F\simeq 0.25\,\rho_6^{2/3}\,{\rm MeV} \ll T$ in  Eq.~\eqref{eq:sigma_fit},
we find the following estimate for Eq.~\eqref{eq:decay_time}
\begin{eqnarray}
  \label{eq:decay_time2a}
  \tau_d \simeq 5\times 10^{11} \left(\frac{\rho}{1~{\rm
      g\,cm^{-3}}}\right)^{0.1} \bigg(\frac{T}{1\,{\rm
      MeV}}\bigg)^{0.8}
  \left(\frac{\lambda_B}{1\,{\rm km}}\right)^2\,{\rm s}.
\end{eqnarray} 
This timescale is clearly much larger than the typical timescales
$\tau_0 \simeq 10\,{\rm ms}$ involved in a merger. Given the current
limitations on the resolution to the order of a meter and choosing the
most favorable temperature and density values we can obtain an
effective lower limit on $\tau_d $ by substituting in
Eq.~\eqref{eq:decay_time2a} $\lambda_B=1\,{\rm m}$,
$\rho \simeq 10^{-3}\,{\rm g\ cm}^{-3}$, and $T=0.1\,{\rm MeV}$, in
which case $\tau_d\sim 10^{4}\,{\rm s}$, which is still
much larger than $\tau_0$ (although smaller rest-mass densities can be
  reached in BNSMs, for $\lambda_B=1\,{\rm m}$ and
  $T=0.1\,{\rm MeV}$ the MHD approach breaks down already at
  $\rho\lesssim 10^{-3}$~g~cm$^{-3}$; see the discussion in
  Sec.~\ref{sec:MHD_val}).  Thus, our analysis suggests that
resistive effects do not play an important role in the MHD
phenomenology of BNSMs and the ideal-MHD approximation in BNSM simulations is always justified as long as the MHD
  description itself is valid. Our conclusion is in contrast with the previous paradigm of
onset of dissipative MHD in low-density regime (Dionysopoulou 2012, Dionysopoulou 2015) where the nearly
zero-conductivity of matter would have implied $\tau_d \to 0$.

\begin{figure} 
\begin{center}
\includegraphics[width=0.487\columnwidth]{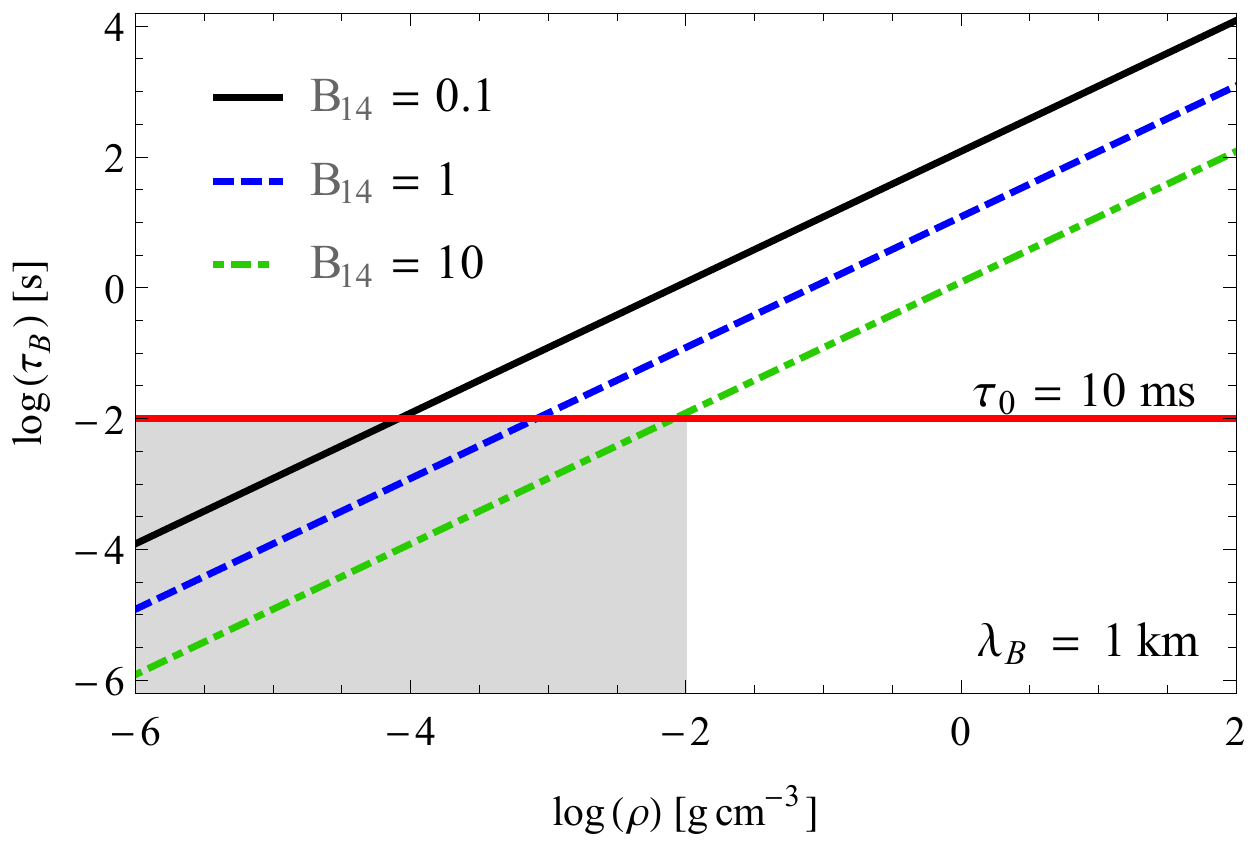}
\hskip 0.2cm
\includegraphics[width=0.488\columnwidth]{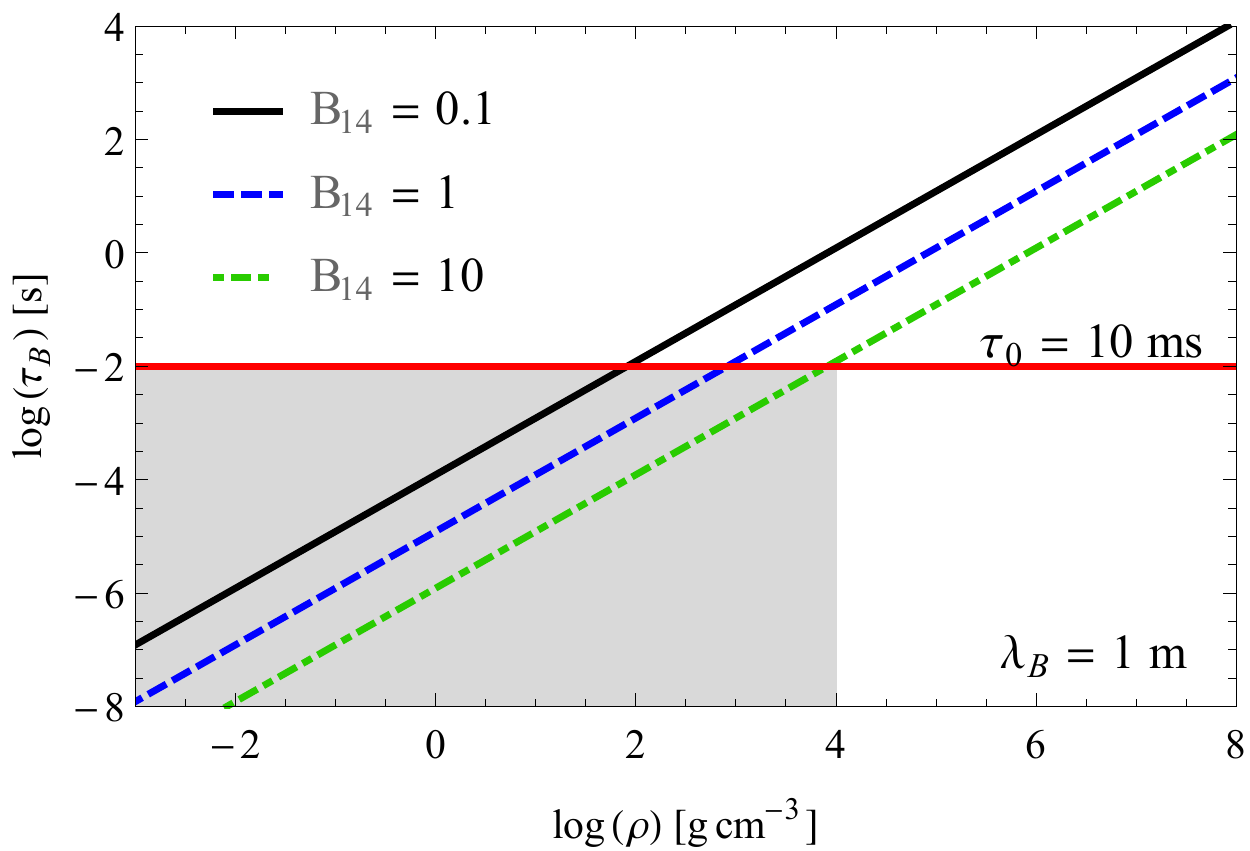}
\caption{Dependence of the Hall timescale $\tau_B$ on the rest-mass
density for two values of the magnetic-field scale-height $\lambda_B=1\,{\rm km}$ 
(\textit{left panel}), and $\lambda_B=1\,{\rm m}$ (\textit{right panel}). 
The solid horizontal lines correspond to the  typical timescale of 
$\tau_0=10\,{\rm ms}$. The shaded areas where $\tau_B\leq\tau_0$
are the regions where the Hall effect becomes important.}
\label{fig:Hall_time_2D}
\end{center}
\end{figure}

In a similar way we express the Hall timescale $\tau_B$ given by Eq.~\eqref{eq:B_time} as 
\begin{eqnarray}
  \label{eq:tau_B1}
  \tau_B \simeq 12\times B_{14}^{-1}
  \left(\frac{\rho}{1~{\rm g\, cm^{-3}}}\right)
  \left(\frac{\lambda_B}{1~{\rm km}}\right)^2\,{\rm s},
\end{eqnarray}
where $B_{14}:=B/(10^{14}\,{\rm G})$. Clearly, the timescale $\tau_B$
is much shorter than the diffusion timescale and it can be in the relevant
range of milliseconds for sufficiently low densities and large magnetic fields. 

The dependence of $\tau_B$ on the rest-mass density is shown in Fig.~\ref{fig:Hall_time_2D} for $\lambda_B=1\,{\rm km}$ and $\lambda_B=1\,{\rm m}$ for several values of the magnetic field. At low densities (the shaded areas in the figure) we have $\tau_B< \tau_0$, therefore in these regions the magnetic field can be subject to the Hall effect, which will change its
distribution on a timescale $\tau_B$. It is seen from Fig.~\ref{fig:Hall_time_2D} that, \eg for  $B_{14}=1$, $\tau_B$
reaches the value $\tau_0=10\,{\rm ms}$ for $\lambda_B=1\,{\rm km}$ at
very low densities $\rho\leq 10^{-3}\, {\rm g\, cm}^{-3}$, while for
$\lambda_B=1\,{\rm m}$ the value $\tau_B = 10\,{\rm ms}$ is reached
already at the densities $\rho\leq 10^{3}\, {\rm g\, cm}^{-3}$, in
agreement with the scaling $\tau_B\propto\rho \lambda^2_B$.

\subsection{Validity of the MHD approach} 
\label{sec:MHD_val}

\begin{figure}[tb] 
\begin{center}
\includegraphics[width=\columnwidth]{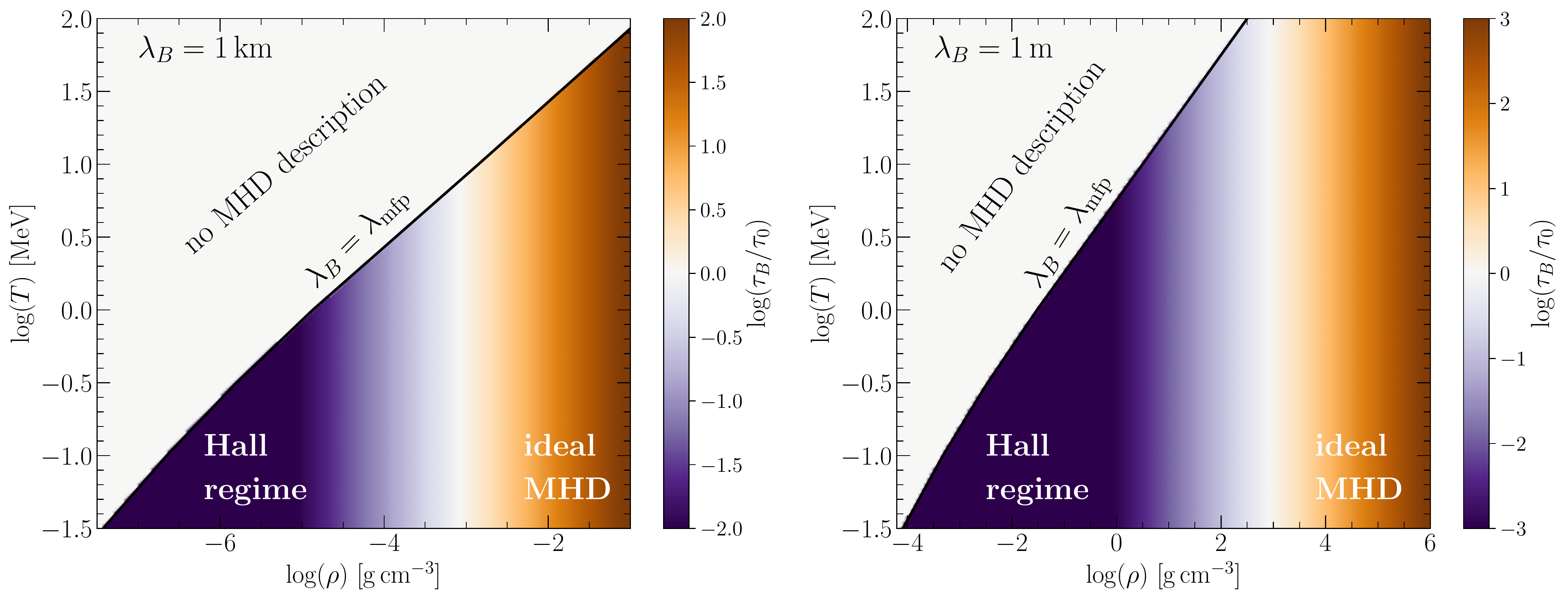}
\caption{Regions of the validity of MHD and ideal MHD on the
  temperature-density plane for  $\lambda_B=1\,{\rm km}$ (\textit{left panel}), and
  $\lambda_B=1\,{\rm m}$ (\textit{right panel}).
   The value of the magnetic field is fixed at
  $B_{14}=1$, and the typical timescale is taken $\tau_0=10\,{\rm
    ms}$. Areas shaded in
  dark-orange are the regions where the ideal-MHD approximation holds;
  areas shaded in dark-violet are the regions where the Hall effect
  becomes important. Above the solid black line $\lambda_B=\lambda_{\rm mfp}$ the MHD
  description of matter breaks down.}
\label{fig:ideal_MHD}
\end{center}
\end{figure}

After having assessed the ranges of validity of ideal MHD, we can now
turn to the next natural question: what limits the applicability of
the MHD approach in the present context?  Clearly, at very low
rest-mass densities the mean free path of electrons $\lambda_{\rm mfp}=\tau v$ becomes large and
the validity of the MHD description of matter itself can break
down. We recall that hydrodynamic description of matter breaks down
whenever $\lambda_{\rm mfp} \geq \lambda_B$.

 At low densities and high temperatures
we can obtain a simple estimate for $\lambda_{\rm mfp}$ [the arguments are similar to those leading to Eq.~\eqref{eq:decay_time2a}]
\begin{eqnarray}
  \label{eq:mfp1}
\lambda_{\rm mfp}\simeq 
4.2\, \left(\frac{\rho}{1~{\rm g\, cm^{-3}}}\right)^{-0.9}
\bigg(\frac{T}{1~{\rm MeV}}\bigg)^{1.8}\,{\rm cm}.
\end{eqnarray}
 Now the condition of applicability of MHD description, \ie $\lambda_{\rm mfp}\le
\lambda_B$, can be written as
\begin{eqnarray}
  \label{eq:MHD_range}
\rho \gtrsim  1.4\times 10^{-5}\, 
\bigg(\frac{T}{1~{\rm MeV}}\bigg)^{2}
\left(\frac{\lambda_B}{1~{\rm km}}\right)^{-1.1}\,{\rm g\, cm^{-3}}.
\end{eqnarray}

It follows from Eq.~\eqref{eq:MHD_range} that the higher the temperature or the smaller the typical lengthscale $\lambda_B$, the higher the rest-mass density below which the
MHD description breaks down.
Figure~\ref{fig:ideal_MHD} shows the regions of validity of the MHD
description in the temperature-density plane for
$\lambda_B=1\,{\rm km}$ and $\lambda_B=1\,{\rm m}$. The low-temperature and high-density region
corresponds to the regime where ideal MHD conditions are
fulfilled. Adjacent to this, a lower density region emerges where the
MHD is applicable, but the Hall effect should be taken into account;
the exact location of this region depends on the strength of the
magnetic field. It moves to lower rest-mass densities for the weaker
magnetic field. Above the separation line
$\lambda_B=\lambda_{\rm mfp}$ the low-density and high-temperature region
features matter in the non-hydrodynamic regime, \ie in a regime where
the MHD approximation breaks down and a kinetic approach based on the
Boltzmann equation is needed.

\section{Summary} 
\label{sec:conclusions}

Motivated by recent developments of multimessenger astronomy started with the observations
of GW170817 in electromagnetic and gravitational waves, we addressed in this work the
role of dissipative processes in the MHD description of BNSMs. Using recently obtained conductivities of warm plasma
in strongly magnetized matter we analysed the timescales for the
evolution of the magnetic field under the conditions which are
relevant for BNSMs. We found that the
magnetic-field decay time is much larger than the relevant timescales
for the merger process in the entire density-temperature range
characteristic for these processes. In other words, the ideal MHD
approximation is applicable throughout the entire processes of the
merger. This conclusion holds for lengthscales down to a meter, which
is at least an order of magnitude smaller than currently feasible
computational grids. Our finding implies a paradigm shift in resistive
MHD treatment of BNSMs, as these were based on
the modeling of conductivities which vanish in the low-density
limit (Dionysopoulou 2015). We have demonstrated that the ideal
MHD description does not break down due to the onset of dissipation,
rather it becomes inapplicable when the MHD description of
matter becomes invalid.

We have demonstrated that the Hall effect plays an important role in
the low-density and low-temperature regime. Thus, the ideal MHD
description must be supplemented by an approach which takes into
account the anisotropy of the fluid via the Hall effect. As is
well-known, the Hall effect can act as a mechanism of rearrangement of
the magnetic field resulting in resistive
instabilities (Gourgouliatos 2016, Kitchatinov 2017). We hope that our
study will stimulate MHD simulations which will include these effects.

\section*{\small Acknowledgements}
\scriptsize{I would like to thank Prof.~Armen Sedrakian, Prof.~Lucciano Rezzolla and Dr.~Antonios Nathanail for scientific collaboration and fruitful discussions. I also acknowledge
support from the HGS-HIRe
graduate program at Goethe University, Frankfurt. The simulations
were performed on the
LOEWE cluster in CSC in Frankfurt by A. Nathanail.}

\end{document}